\documentclass[useAMS]{mn2e}
\usepackage{graphicx}
\title[On the universality of density profiles]{On the universality of density profiles}
\author[A. Del Popolo]{A. Del Popolo$^{1, 2}$\thanks{E-mail:
antonino.delpopolo@unibg.it} \\
$^{1}$Dipartimento di Fisica e Astronomia, Universit\'a di Catania, Viale Andrea Doria 6, 95125 Catania, Italy\\
$^{2}$Argelander-Institut f\"ur Astronomie, Auf dem H\"ugel 71, D-53121 Bonn, Germany}
\begin{document}


\pagerange{\pageref{firstpage}--\pageref{lastpage}} \pubyear{2002}

\maketitle

\label{firstpage}

\begin{abstract}
We use the secondary infall model described in Del Popolo (2009),
which takes into account the effect of dynamical friction, 
ordered and random angular momentum, baryons adiabatic contraction and dark matter baryons interplay, to study how 
inner slopes of relaxed $\Lambda$CDM dark matter (DM) halos with and without baryons (baryons+DM, and pure DM)
depend on redshift and on halo mass. We apply the quoted method to 
structures on galactic scales and clusters of galaxies scales. We find that the inner logarithmic density slope, $\alpha\equiv d\log\rho/d\log r$, of dark matter halos with baryons has a significant dependence on halo mass and redshift with slopes ranging from $\alpha \simeq 0$ for dwarf galaxies to $\alpha \simeq 0.4$ for objects of 
$M \simeq 10^{13} M_{\odot}$ and $\alpha \simeq 0.94$ for $M \simeq 10^{15} M_{\odot}$ clusters of galaxies. Structures slopes increase with increasing redshift 
and this trend reduces going from galaxies to clusters. 
In the case of density profiles constituted just of dark matter the mass and redshift dependence of slope is very slight. In this last case, we used the Merrit et al. (2006) analysis who compared $N$-body density profiles with various parametric models finding systematic variation in profile shape with halo mass. This last analysis suggests that the galaxy-sized halos obtained with our model have a different shape parameter, i.e. a different mass 
distribution, than the cluster-sized halos, obtained with the same model. The results of the present paper argue against universality of density profiles constituted 
by dark matter and baryons and confirm claims of a systematic variation in profile shape with halo mass, for dark matter halos.
\end{abstract}

\begin{keywords}
cosmology--theory--large scale structure of Universe--galaxies--formation
\end{keywords}

\section{Introduction}

A fundamental idea that has come out of the numerical approach is that relaxed halos are (nearly) universal in many respects: 
the phase-space density profile (Taylor \& Navarro 2001), the linear relation between the density slope and the velocity anisotropy (Hansen \& Moore 2006), the density profile (Navarro, Frenk \& White 1996 (hereafter NFW)), the distribution of axial ratio, the distribution of spin parameter, and the distribution of internal
specific angular momentum (Wang \& White 2008). 

For what concerns the density profiles, numerous cosmological studies (e.g., Dubinski \& Carlberg 1991; Crone at al. 1994; Navarro et al. 1996, 1997; 
Carlberg et al. 1997; Moore et al. 1998, 1999; Jing \& Suto 2000)
revealed similar density profiles over several
orders of magnitude in halo mass, with a central cusp and a
$\rho(r) \propto r^{-3}$ fall--off at large radii. 
The universal shape of the profiles, found by Navarro et al. (1996, 1997), has been confirmed by several other studies, even if the actual value of the inner density slope 
$\alpha$ has been a matter of controversy (Moore et al. 1998; Jing \& Suto 2000; Ghigna et al. 2000; Fukushige \& Makino 2001) and more recently, the 
functional form 
of the universal profile has been substituted by profiles whose logarithmic slope becomes increasingly shallower inwards (Power et al. 2003; Hayashi et al. 2003 and Fukushige et al. 2004; Navarro et al. 2004; Stadel et al. 2008). Moreover, there is also discussion whether the inner slope is actually universal or not 
(Moore et al. 1998; Jing \& Suto 2000; Subramanian et al. 2000; Klypin et al. 2001; Ricotti 2003, Ricotti \& Wilkinson 2004; Cen et al. 2004; Navarro et al. 2004; Fukushige et al. 2004; Merrit et al. 2005; Merrit et al. 2006; Graham et al. 2006; Ricotti et al. 2007; Gao et al. 2008; Host \& Hansen 2009).

All of the previously quoted analyses dealing with universality of dark matter profiles do not study the possible effects produced by the presence of baryons, whose effect is to shallow (El-Zant et al. 2001, 2004; Romano-Diaz et al. 2008) and to steepen (Blumenthal et al. 1986; Gnedin et al. 2004; Klypin et al. 2002) the dark matter profile.
In real galaxies, the two quoted effects combine, with the result of giving rise to density profiles which are different from those predicted in N-body simulations (Del Popolo 2009).
In collisionless N-body simulations, this complicated interplay between baryons and dark matter
is not taken into account, because it is very hard to include the effects of baryons in the simulations. However, in order to have a clear view of what simulations can tell about the inner parts of the density profiles, it is necessary to run N-body simulations that repeat the mass modeling including a self-consistent treatment of the baryons and dark matter component. The question is whether or not baryon-DM interactions are universal, or depend on scale - that could be answered by simulations in the next few years.
%
%

In the present paper, we deal with the problem of the universality of density profiles made of dark matter and baryons, by using the results of Del Popolo (2009).

In the next sections, we use Del Popolo (2009) model to study how inner slopes in density profiles change when baryons are present. The paper is organized as follows: in Section 2, we describe the Del Popolo (2009) model. In Section 3, we discuss the results. Finally, Section 4 is devoted to conclusions.

\section[]{Model}

The dark matter halos in the present study are formed using the analytical method introduced in Del Popolo (2009).
In this section, we shortly summarize the model that will be used to study the behavior of the inner part of dark matter haloes when baryons are present.

In the secondary infall model (SIM) (Gunn \& Gott 1972), 
a bound mass shell of initial comoving radius $x_i$ will expand to a maximum radius $x_m$ (named apapsis or turnaround radius $x_{ta}$).
As successive shells expand to their maximum radius, they acquire angular momentum and then contract on orbits determined by the angular momentum. Dissipative physics and the process of violent relaxation will eventually intervene and convert the kinetic energy of collapse into random motions (virialization). 


The final density profile can be obtained in terms of the density at turn-around, $\rho_{ta}(x_m)$, the collapse factor\footnote{The collapse factor is defined as the $f=x/x_m$ (see Del Popolo 2009, Appendix A)}, and the turn-around radius (Eq. A18, Del Popolo 2009),
as:
\begin{equation}
\rho(x)=\frac{\rho_{ta}(x_m)}{f^3} \left[1+\frac{d \ln f}{d \ln x_m} \right]^{-1}
\label{eq:dturnnn}
\end{equation}
In our model we supposed that protostructure forms around peaks of the density field, we took into account the presence of baryons, baryons adiabatic collapse, dynamical friction, and angular momentum. These quantities can be specified as we describe in the following.

The density profile of a proto-halo is taken to be the profile of a peak in a density field described by the Bardeen et al. (1986) power spectrum, as is illustrated in Del Popolo (2009), Figure 6.

In the present paper, we take into account the ordered angular momentum, $h$, (Ryden \& Gunn (hereafter RG87)) which arises from tidal torques experienced by proto-halos, 
and the random angular momentum, $j$, (RG87)) which is connected to random velocities (see RG87).
Ordered angular momentum is got obtaining, first, the rms torque, $\tau (x)$, on a mass shell,
then obtaining total specific angular momentum, $h(r,\nu )$, acquired during expansion by integrating the torque over time (Ryden 1988a (hereafter R88), Eq. 35) (see Appendix C of Del Popolo 2009).

The random part of angular momentum was assigned to protostructures according to Avila-Reese et al. (1998) scheme. This consists in expressing the specific angular momentum $j$ through the ratio $e_0=\left( \frac{r_{min}}{r_{max}} \right)_0$, where $r_{min}$ and $r_{max}$ are the maximum and minimum penetration of the shell toward the center, respectively,
and left this quantity as a free parameter (see Appendix C of Del Popolo 2009). 
 
In the present paper, we took into account dynamical friction by introducing the dynamical friction force in the equation of motion (see Del Popolo 2009, Eq. A16). Dynamical friction force was calculated dividing the gravitational field into an average and a random component generated by the clumps constituting hierarchical universes following Kandrup (1980) (see Appendix D of Del Popolo 2009).

The shape of the central density profile is influenced by baryonic collapse: baryons drag dark matter in the so called 
adiabatic contraction (AC) steepening the dark matter density slope.
%
%
The adiabatic contraction was taken into account by means of Gnedin et al. (2004) model and Klypin et al. (2002) model taking also account of exchange of angular momentum between baryons and dark matter (see appendix E of Del Popolo 2009 for a wider description). 
Our method of halo formation has considerable flexibility with direct control over the parameter space of initial conditions differently from 
numerical simulations which yield little physical insight beyond empirical findings precisely because 
they are so rich in dynamical processes, which are hard to disentangle and 
interpret in terms of underlying physics.

\section{Results and discussion}

\subsection{Mass dependence of the inner slope}

After describing the main points of the model, fully described in Del Popolo (2009), we will use it to determine the inner density slope of density profiles generated by the model. We calculate the logarithmic slopes, $\alpha\equiv d\log\rho/d\log r$, of the $\Lambda$CDM halos for different value of the mass
at the radius $\sim 10^{-2}r_{vir}$ for three different cases: in the case A we take into account all the effects included in Del Popolo (2009), namely angular momentum, dynamical friction, baryons, baryons adiabatic contraction; in the case B 
there are no baryons and dynamical friction, and angular momentum is taken into account as in Del Popolo (2009); in case C only angular momentum is taken into account 
reduced, as in Del Popolo (2009), in order to reproduce the angular momentum of N-body simulations (dashed line in Fig. 3) and a NFW profile (solid histogram in Fig. 3).
We recall that in Del Popolo (2009), we performed an experiment similar to that performed by Williams et al. (2004) namely we reduced the magnitude of the $h$ and $j$ angular momentum, of a factor of 2, and dynamical friction force of a factor 2.5 with respect to the 
typical values calculated and used in the model in order to reproduce angular momentum of N-body simulations and the NFW profile. The density profiles in Del Popolo (2009) (as in Williams et al. 2004) are shallower than N-body profiles. By reducing the amplitude of angular momentum, the profiles get steeper at the center and one can find a value of angular momentum 
for which the NFW profile is reproduced, corresponding to the angular momentum seen in N-body simulations (solid line in Fig. 3). This effect can be understood, as already reported: the central density is built up by shells whose pericenters are very close to the center of the halo. Particles with larger angular momenta are prevented from coming close to the halo's center and so contributing to the central density. 
The correlation between increasing angular momentum and the reduction of inner slopes in halos has been also noticed by several other authors (Avila-Reese et al. 1998, 2001; Subramanian et al. 2000; Nusser 2001; Hiotelis 2002; Le Delliou \& Henriksen 2003; Ascasibar et al. 2003). 

\begin{figure}
(a)
\includegraphics[width=84mm]{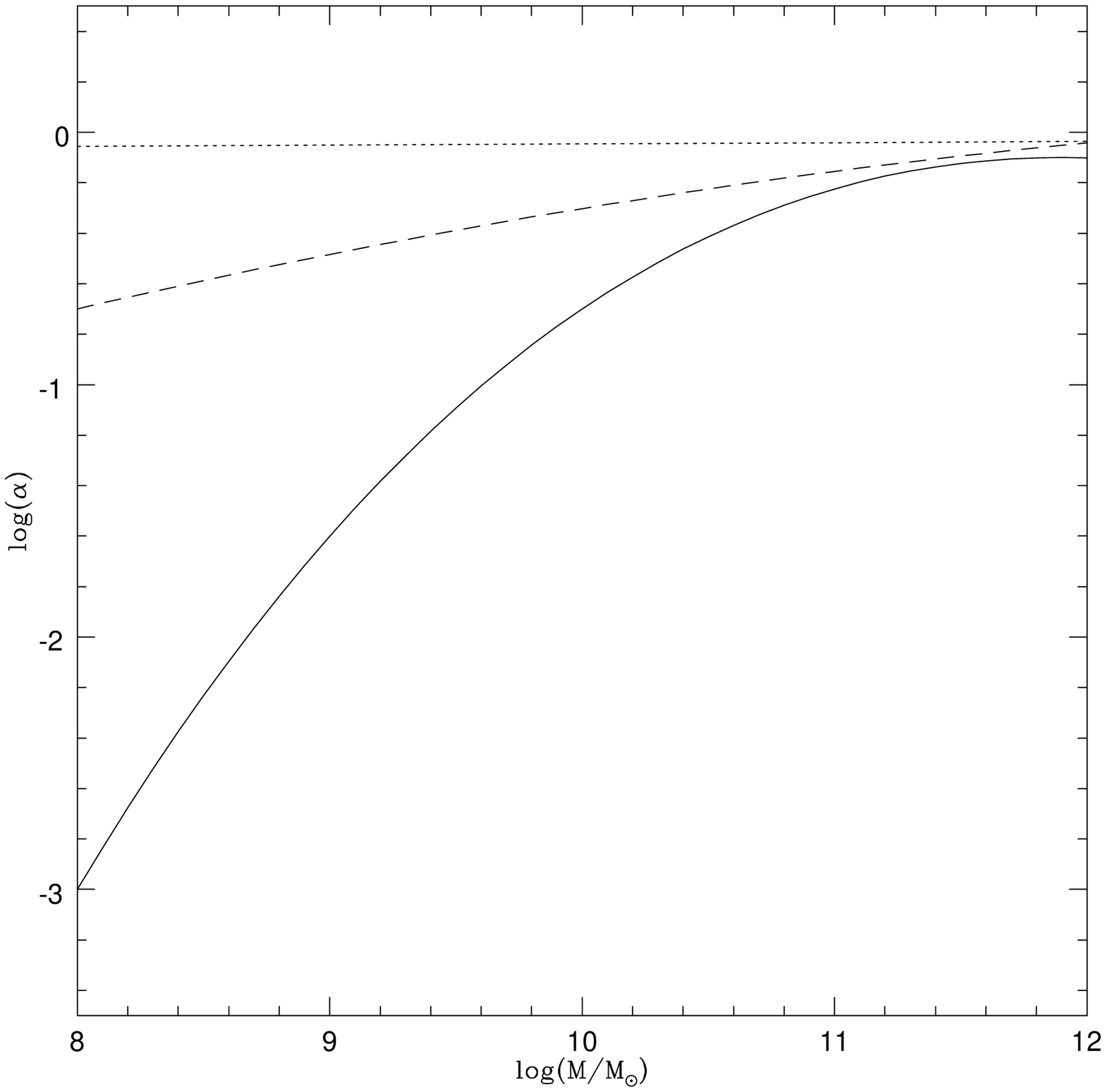} (b)
\includegraphics[width=84mm]{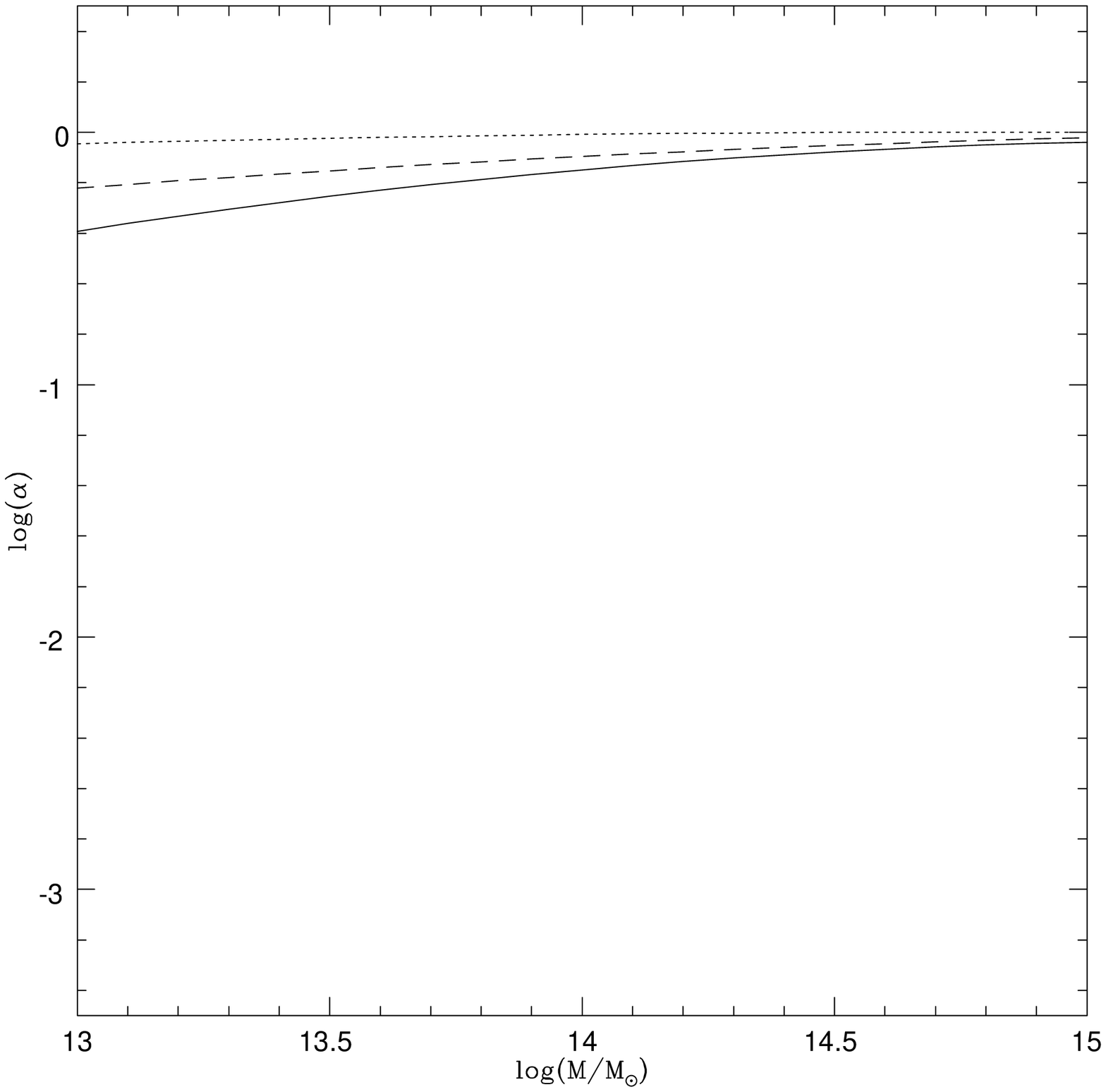} 
\vspace{3.5cm}
\caption{
Inner density slope as a function of halo mass. Panel (a): Solid line, dashed line, and dotted line represents the result of the model of the present paper for the case A, B, and C described in the text. Panel (b): same as panel (a) but for groups and clusters of galaxies.
}
\end{figure}

\begin{figure}
\hspace{-1.0cm} (a)
\includegraphics[width=84mm]{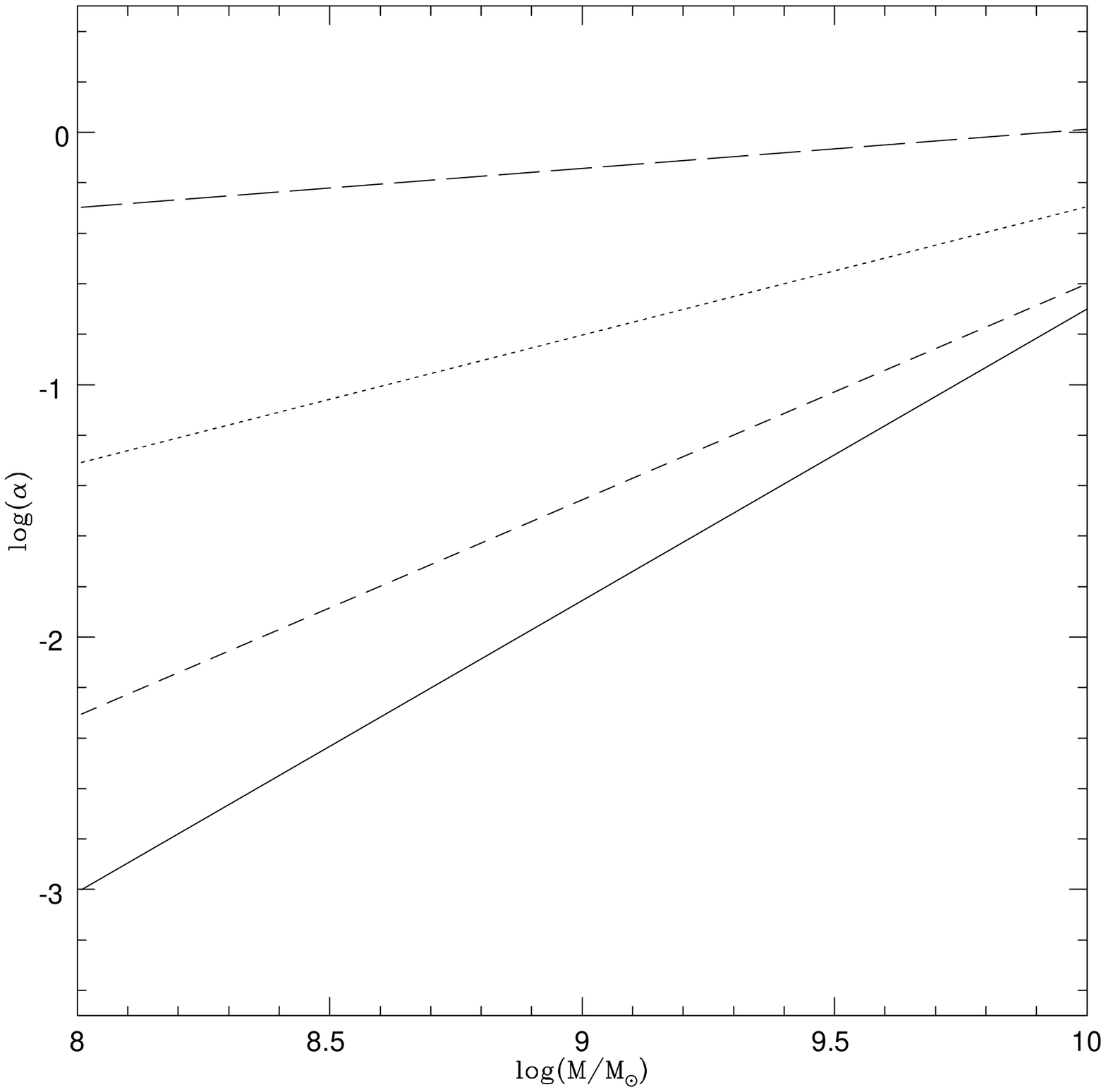} 
\hspace{-2.0cm} \includegraphics[width=84mm]{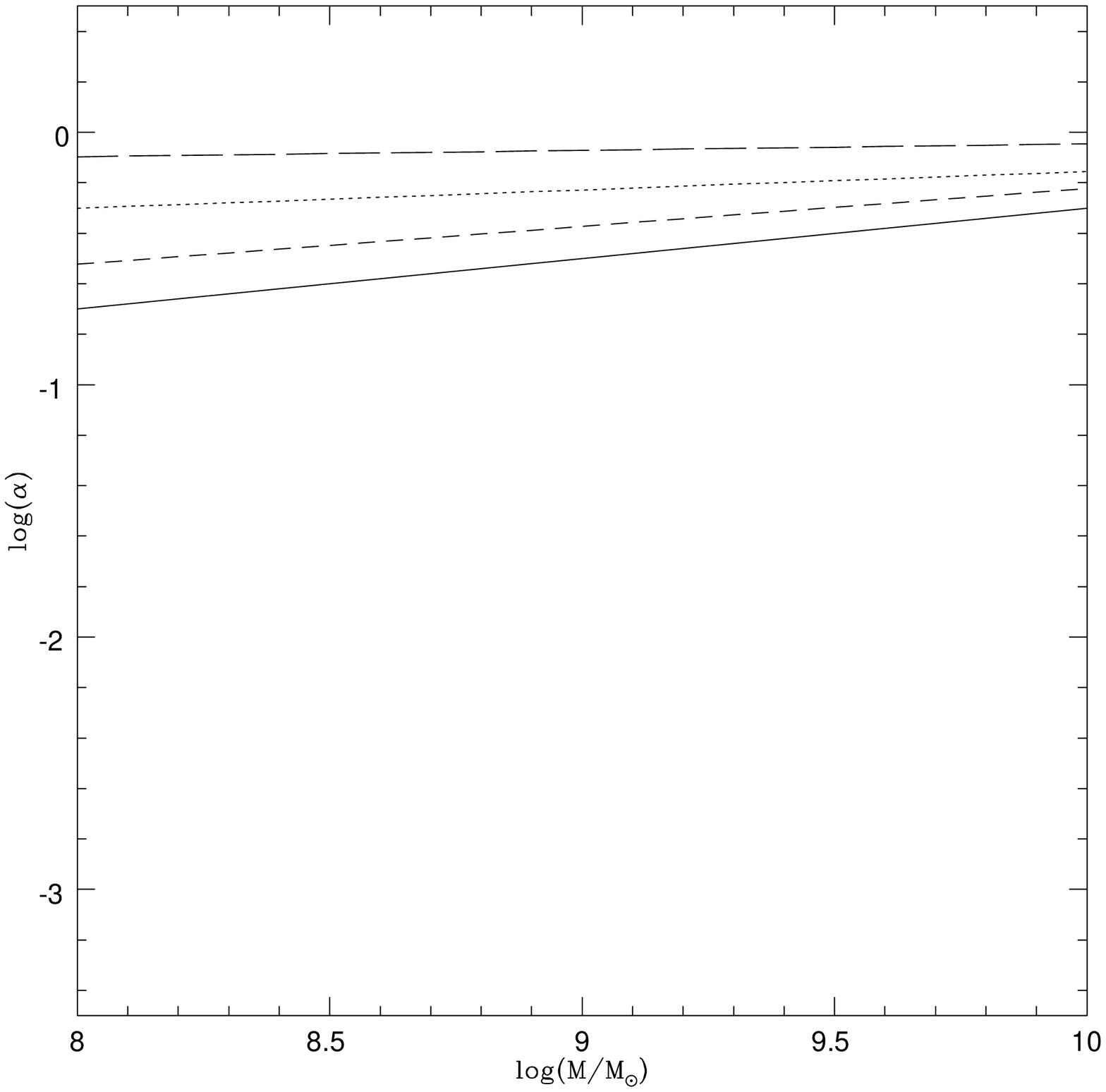} (b)
%
\caption{
Inner density slope as a function of halo mass and redshift. Panel (a): solid line, long-dashed line, dotted line, and dashed line, represent the redshifts $z=0, 1,2,2.5$, respectively, for case A. Panel (b): same as panel (a) but for case B. Panel (c), and Panel (d), same as panel (a) and (b) but for clusters of galaxies and for redshift
$z=0,1,2$, solid, dashed and dotted lines, respectively. 
}
\end{figure}

\begin{figure}
\hspace{-1.0cm} (c)
\includegraphics[width=84mm]{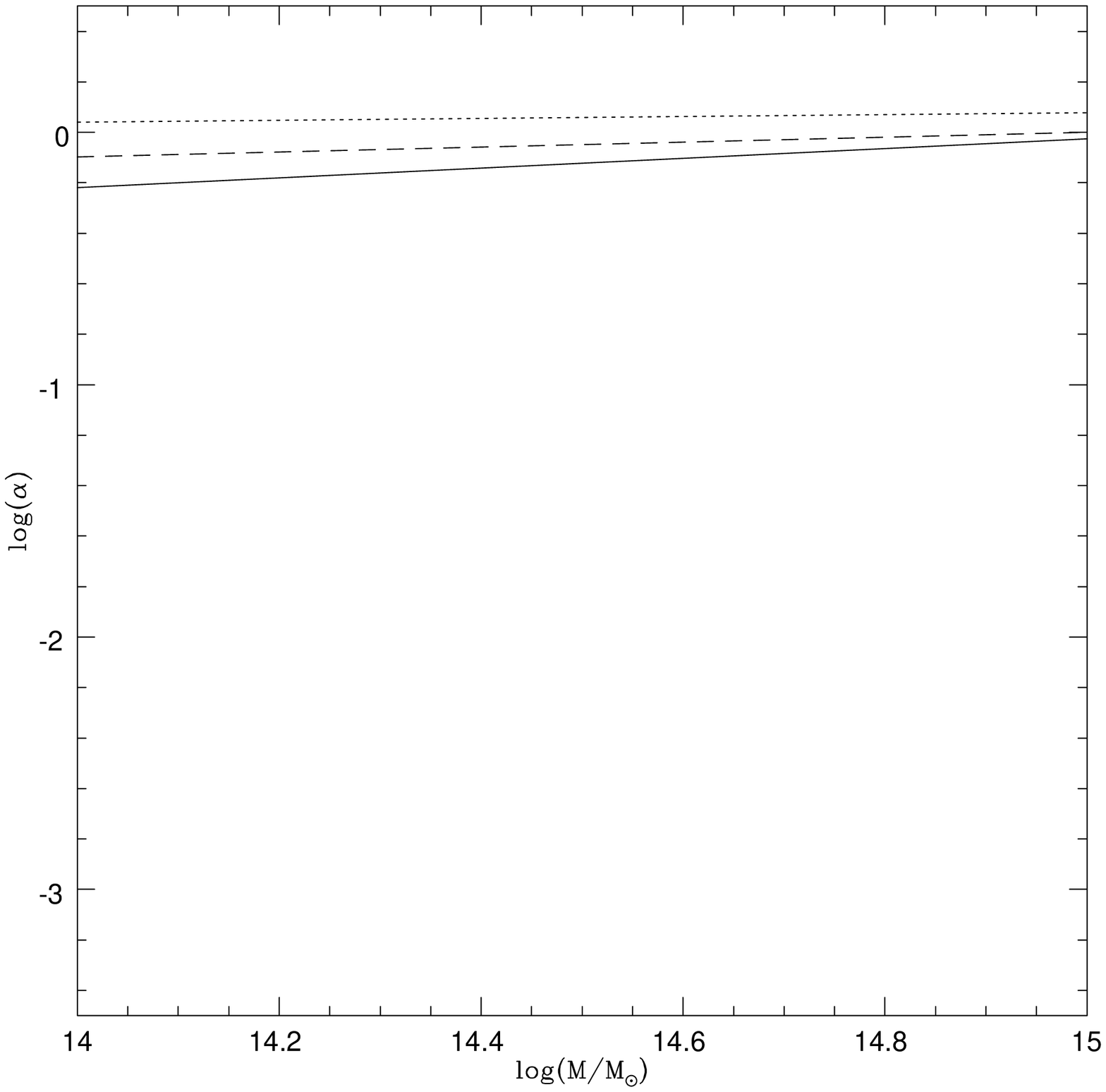} 
\includegraphics[width=84mm]{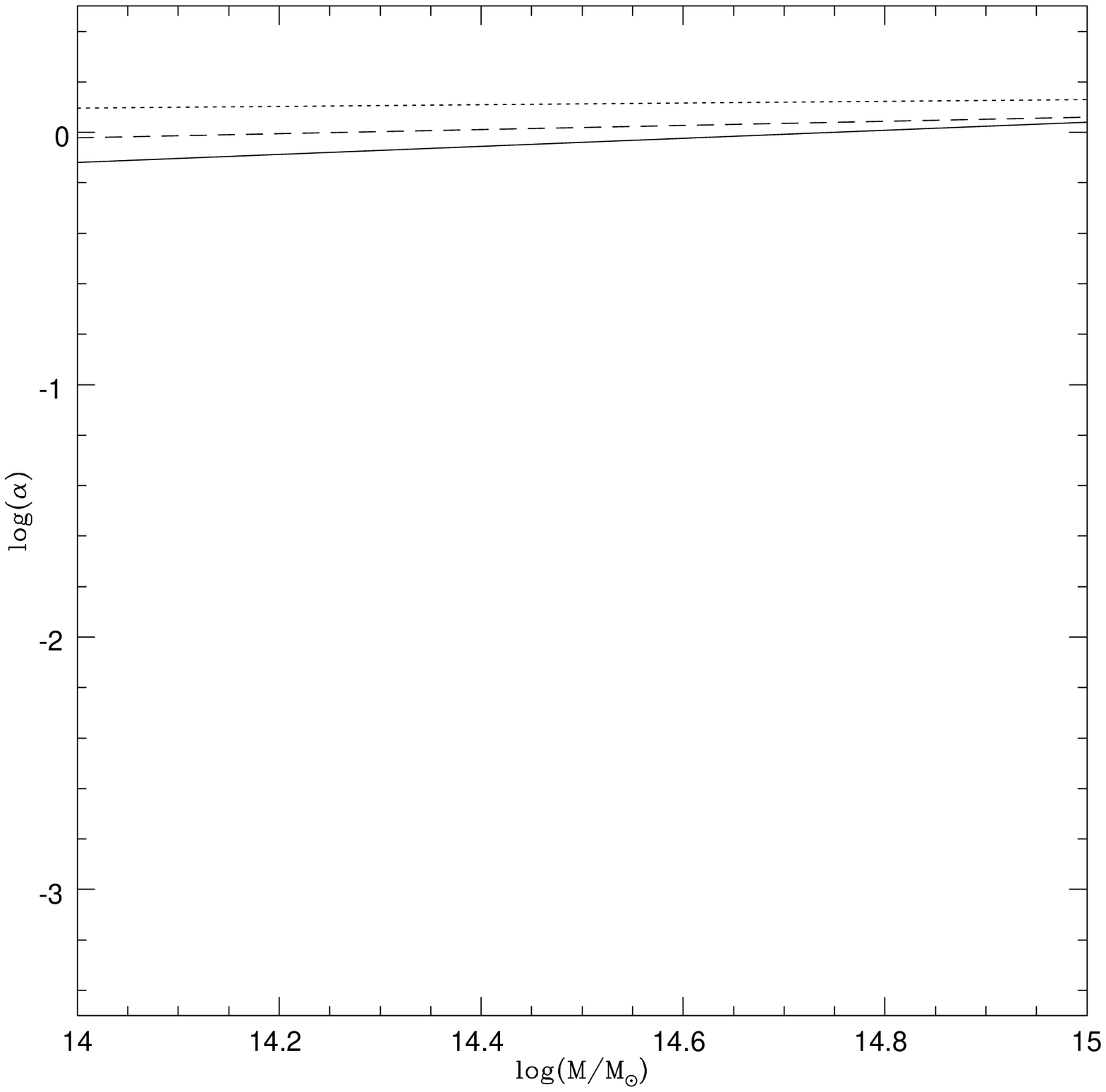} (d)
\end{figure}


In Fig. 1, we plot $\alpha$, at $z=0$, in terms of the mass for galaxies in the mass-range $10^8-10^{12} M_{\odot}$. The solid, dashed and dotted lines represents, respectively: (a) the slope in the case A; 
b) the slope in the case B;
(c) the slope in the case C.
The solid line (result of the case A) shows that the inner profiles of dwarf galaxies are very flat with
logarithmic slope $\alpha \simeq 0$ for $10^8-10^{9} M_{\odot}$. 
Our results shows a steepening of the density profile with increasing mass with slopes 
$\alpha \simeq 0.2$ for $M \simeq 10^{10} M_{\odot}$, $\alpha \simeq 0.6$ for $M \simeq 10^{11} M_{\odot}$, 
$\alpha \simeq 0.8$ for $M \simeq 10^{12} M_{\odot}$. If we do not take into account baryons and dynamical friction (case B) (dashed line), the shape of the density profiles becomes steeper, with respect to case A, and the inner slopes are $\alpha \simeq 0.2$ for $M \simeq 10^{8} M_{\odot}$, $\alpha \simeq 0.33$ for $M \simeq 10^{9} M_{\odot}$, $\alpha \simeq 0.5$ for $M \simeq 10^{10} M_{\odot}$, $\alpha \simeq 0.7$ for $M \simeq 10^{11} M_{\odot}$, and $\alpha \simeq 0.91$ for $M \simeq 10^{12} M_{\odot}$. In the last case, dotted line represents the quoted case C, 
and the profiles are even steeper than the previous case B and the inner slope are $\alpha \simeq 0.88$ for $M \simeq 10^{8} M_{\odot}$, $\alpha \simeq 0.90$ for $M \simeq 10^{9} M_{\odot}$, $\alpha \simeq 0.91$ for $M \simeq 10^{10} M_{\odot}$, $\alpha \simeq 0.92$ for $M \simeq 10^{11} M_{\odot}$, and $\alpha \simeq 0.93$ for $M \simeq 10^{12} M_{\odot}$. 
The differences in slopes with mass for the three plotted cases (A, B, and C) can be explained as follows. In case A, represented by the solid line, baryons, dynamical friction and angular momentum are present. The final density profile and final slope is determined by the interplay of these three factors. Let's see how each of these factors act in shaping the profile and how they interplay. 
For what concerns angular momentum, we have to note that less massive objects are generated by peaks of smaller $\nu$, which acquire more angular momentum ($h$ and $j$).
The angular momentum sets the shape of the density profile at the inner regions. For pure radial orbits, the core is dominated by particles from the outer shells. As the angular momentum increases, these particles remains closer to the maximum radius, resulting in a shallower density profile. Particles with smaller angular momentum will be able to enter the core but with a reduced radial velocity compared with the purely radial SIM. For some particles, the angular momentum is so large that they will never
become unbound. Summarizing, particles of larger angular momenta are prevented from coming close to the halo's center and
so contributing to the central density, which has the effect of flattening the density profile. 
%
%
The effects of dynamical friction can be interpreted in two different fashions: (a) an increase in dynamical friction force
is very similar to changing the magnitude of angular momentum (see Fig. 11 of Del Popolo 2009) with the final result of producing shallower profiles; (b) dynamical friction can act on gas moving in the background of dark matter particles, dissipate the clumps orbital energy and deposit it in the dark matter with the final effect of erasing the cusp (similarly to El-Zant et al. 2001 (hereafter EZ01); El-Zant et al. 2004; Tonini, Lapi \& Salucci 2006 (hereafter TLS); Romano-Diaz et al. 2008). Baryons have another effect, at an early redshift, the dark matter density experiences the adiabatic contraction by baryons producing a slightly more cuspy profile. This last is overcome from the previous two effects. As shown by Fig. 11 of Del Popolo (2009), the magnitude of dynamical friction effect is a bit larger than that due to angular momentum and that these two effects add to improve the flattening of the profile.
The quoted effects act in a complicated interplay. Initially, at high redshift (e.g., $z=50$), the density profile is in the linear regime. The profile evolves to
the non-linear regime, and virialize.
At an early redshift, (e.g. $z \simeq 5$, for dwarf galaxies), the dark matter density experiences the adiabatic contraction by baryons producing a slightly more cuspy profile.
The evolution after virialization is produced by secondary infall, two-body relaxation, dynamical friction and angular momentum. 
Angular momentum, as described, contributes to reduce the inner slope of density profiles by preventing particles from reaching halo's center, while dynamical friction 
dissipate the clumps orbital energy and deposit it in the dark matter with the final effect of erasing the cusp (similarly to EZ01; El-Zant et al. 2004; TLS; Romano-Diaz et al. 2008).
The cusp is slowly eliminated and within $ \simeq 1$ kpc a core forms, for objects of the mass of dwarf-galaxies. 

It is now clear why going from a model which takes into account baryons, dynamical friction, and angular momentum (solid line, case A) to one taking account just of angular momentum (dashed line, case B) and to one taking account just of angular momentum reduced to reobtain N-body simulations angular momentum,
one obtains larger inner slopes.

In Fig. 1b, we plot the logarithmic slope in the case of groups and clusters of galaxies similarly to Fig. 1a. The change of slope with mass is qualitatively similar to the case of galaxies. However, in the case of groups and clusters the profiles are steeper showing a larger value of $\alpha$. 
We obtain the following results: for masses $M \simeq 10^{13} M_{\odot}$, $M \simeq 10^{14} M_{\odot}$, $M \simeq 10^{15} M_{\odot}$ we have
$\alpha \simeq 0.4, 0.6, 0.94$, respectively for case A. For case B the slopes are $\alpha \simeq 0.6, 0.8, 0.95$, and for case C, $\alpha \simeq 0.96, 0.98, 1$.

The main reason of the difference of behavior between groups-clusters and galaxies is due to the fact that in the case of clusters the virialization process starts much later with respect to galaxies. In the case of galaxies the profile strongly evolve after virialization through the processes previously described. In the case of dwarf galaxies of $10^9 M _{\odot}$, we showed in Del Popolo (2009) that the profile virializes at $z \simeq 10$ and from this redshit to $z=0$ its shape continues to evolve. In the case of a cluster of $10^{14} M _{\odot}$, 
the profile virializes at $z \simeq 0$ (Del Popolo 2009) and, as a consequence, the further evolution observed in galaxies cannot be observed in clusters. 

\subsection{Mass and redshift dependence of the inner slope}

In Fig. 2a, we study the change of 
slope with redshift for dwarf galaxies in the case A.
The solid line, long-dashed line, dotted line, and dashed line, represents the evolution of $\alpha$ with mass and for redshifts $z=0, 1,2, 2.5$, respectively. 
The evolution with mass of $\alpha$ was clearly shown in Fig. 1, while Fig. 2 shows that for larger redshifts there is an increase of $\alpha$, in the indicated range of redshifts. This behavior is expected because as shown in Del Popolo (2009) (e.g. their Fig. 3) the profiles steepens and $\alpha$ is larger for higher values of redshift. In fact as we already reported, in the case of a $10^9 M _{\odot}$ galaxy (Fig. 3 of Del Popolo 2009) from virialization ($z=10$) to $z=0$ there is a continuous decrease in the slope, except at $z \simeq 5$ when adiabatic contraction steepen the profile. For dwarfs of mass larger than $10^{10} M _{\odot}$ the redshift of virialization and the redshift at which adiabatic contraction steepens and the profile are a bit smaller than for $M \simeq 10^{9} M_{\odot}$. Fig. 2b shows how the slope depends on redshift for the model B.

The quoted figure shows that the change of the slope with redshift is less strong than of case A. This is due to the fact that when baryons and dynamical friction are taken into account the profile evolution is faster. The model corresponding to the case C has slopes that
almost has no dependence on redshift, and for this reason we did not plot it. 
Fig. 2c, and Fig. 2d is the same as Fig 2a, Fig. 2b but for clusters. We observe a less strong dependence of the slope on redshift, with respect to galaxies, for the reason already described.

The clear result of Fig. 1 and Fig. 2 is that the inner slope of density profiles of dark matter and baryons (case A) and density profiles of dark matter characterized by standard angular momentum (case B) depends on mass and on redshift. This result argues against the universality of this kind of profiles.

\begin{figure}
\includegraphics[width=84mm]{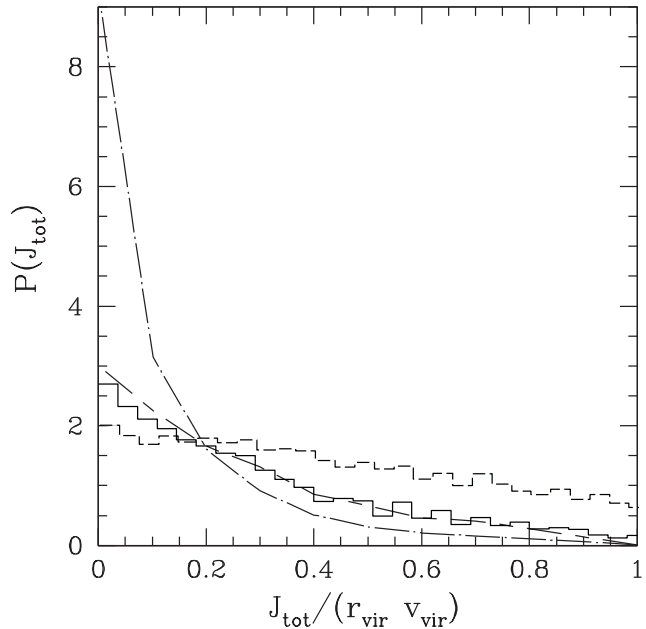} 
\caption[]{Distribution of the total specific angular momentum, $J_{Tot}$. The dotted-dashed and dashed line represents the quoted 
distribution for the halo n. 170 and n. 081, respectively, of van den Bosch et al. (2002). The dashed histogram is the distribution obtained 
from our model for the $10^{12} M_{\odot}$ halo and the solid one the angular momentum distribution for the density profile reproducing the NFW halo, as descried in the text.
}
\end{figure}

As previously reported, the dependence of the inner slope with mass has been observed in several N-body simulations (e.g. Navarro et al. 2004) and has received different interpretations. Gao et al. (2008) used two very large cosmological simulations to study how the density profiles of relaxed $\Lambda$CDM dark halos depend on redshift and on halo mass. The profiles by them obtained deviate slightly but systematically from
the NFW form and are better approximated by an Einasto profile. They found that the shape parameter of the quoted profile changes with mass and redshift (see their Fig. 2). However, 
they interpret the result as 
supporting the idea that halo densities reflect the density of the universe at the time they formed, as proposed by Navarro, Frenk \& White
(1997). From this point of view, the shape parameter, $\alpha$, should be used to improve the description of the typical density profiles of simulated
halos and to eliminate possible biases in estimates of their concentration. Similarly Navarro et al. (2004), and Klypin et al. (2001) 
interpret the change in the value of the slope as a reflection of the trend between the concentration of a halo and its mass. 
Other studies, namely Merrit et al. (2005, 2006) interpret the systematic variation in profile shape with halo mass, as indicating that $\Lambda$CDM halos have not 
a truly universal profile, in agreement with previous quoted studies (Jing \& Suto 2000; Subramanian et al. 2000; Ricotti 2003, Ricotti \& Wilkinson 2004; Cen et al. 2004; Merrit et al. 2005; Ricotti et al. 2007; Host \& Hansen 2009). 

\begin{figure}
\includegraphics[width=84mm]{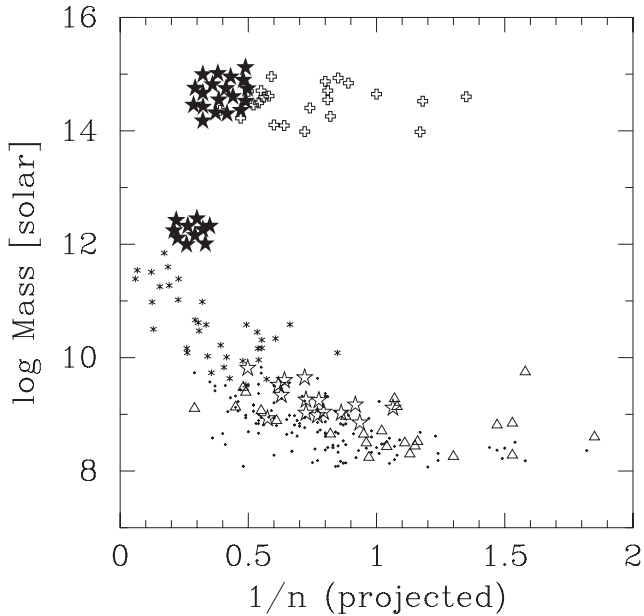} 
\caption{
Mass versus profile shape ($1/n$). 
Dark matter halos are obtained with the model of the present paper, case C.
We plot the virial masses and the shape parameters calculated from  the best-fitting
Prugniel-Simien model.
For the galaxies and galaxy clusters, the shape parameters $n$ have come from
the best-fitting S\'ersic 
model to the (projected) luminosity-and X-ray profiles, respectively. We used Merrit et al. (2006) results.
In the case of dark matter halos, we used the halos obtained through Del Popolo(2009) and calculated the best-fitting Prugniel-Simien model.
%
%
Filled stars: halos from the present paper, case C;
open plus signs: galaxy clusters from Demarco et al.\ (2003);
dots: dwarf Elliptical (dE) galaxies from Binggeli \& Jerjen (1998);
triangles: dE galaxies from Stiavelli et al.\ (2001);
open stars: dE galaxies from Graham \& Guzm\'an (2003); 
asterisk: intermediate to bright elliptical galaxies from Caon et al.\ 
(1993) and D'Onofrio et al.\ (1994). 
}
\end{figure}

We should stress that in the case of DM N-body simulations the deviations from the NFW are slight, and for this reason one can arrive to different interpretations 
and to try to reduce the discrepancy by modifying the definition of halo formation time, the concentration-mass relation, as in Gao et al. (2008).\footnote{Even if with the quoted interpretation, the  evolution of the concentrations of Milky Way mass halos is still not reproduced well.}
In our case A, density profiles have a final shape highly different from that of a NFW density profile or its improvements (e.g., Navarro et al. 2004), since our profiles are not containing just DM but also baryons. For this reason, a comparison of our result for the case A with previous ones, is not easy since all previous studies finding or not, a universal behavior in density profiles, are  dealing just with DM halos: they do not take account of the effect of baryons. 
In the case A of our model there is no room for different interpretations as in previous studies: the density profiles are so different from those obtained in N-body simulations, and the inner slopes changes so strongly with mass and redshift that we are compelled to the conclusion that when baryons are taken into account the profiles are not universal.   
\begin{figure}
\includegraphics[width=84mm]{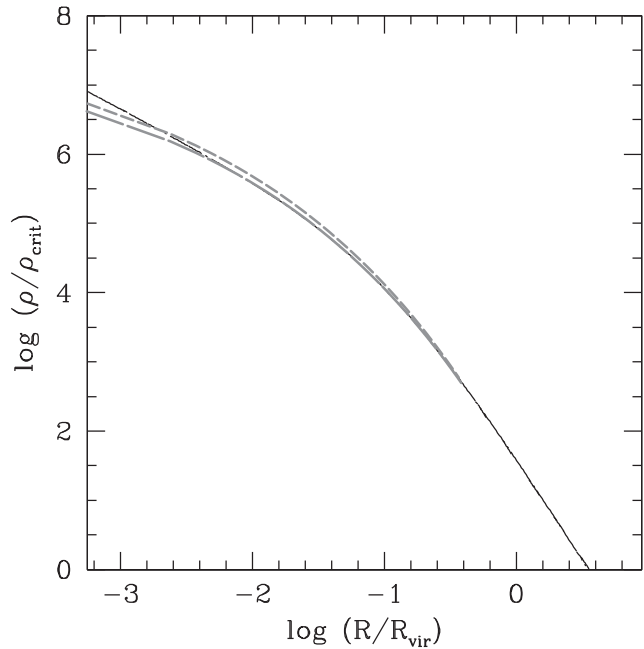} 
\caption{Comparison of halos generated by the model of the present paper (long dashed--line) with numerical simulations. The solid line is the NFW profile while the short--dashed line result of the Aquarius experiment, both for an halo of $10^{12} M_{\odot}$.}
\end{figure}

Similar conclusions are valid for the case B: the mass and redshift dependence of the inner slope, $\alpha(M,z)$, is so strong that again we are lead to infer that the profile in this case are not universal. 

In case C, as discussed the mass and redshift dependence of the slope, are quite slight. This means that in the case of a DM proto-structure with angular momentum reduced to reproduce angular momentum in N-body simulations (dashed line or solid histogram in Fig. 3) 
the density profiles and slopes have a similar behavior to those obtained in N-body simulations. 
In this third case, the slight slope change with mass could be interpreted as in Merrit et al. (2006) or differently as in Navarro et al. (2004), and Klypin et al. (2001). 
However, even following the interpretation of
Navarro et al. (2004), who wrote ``adjusting the parameter $n$ allows the profile to be tailored to each individual halo, resulting in improved fits''\footnote{The value of $n$, equal to $1/\alpha$ in Navarro et al.'s (2004) notation, ranged from 4.6 to 8.2 (Navarro et al. 2004, their table~3).}, such a breaking of structural homology (see Graham \& Colless 1997 for an analogy with projected
luminosity profiles) replaces the notion that a universal density profile may exist.

Before concluding the present subsection, we recall that the model of the present paper gives just one profile for a given halo mass. In reality N-body simulations predict an ensemble of profiles for a given halo mass, and that slope vary from halo to halo (halo--to--halo scatter) (Reed et al. 2005), depending on the different formation histories. 
%
%
In particular, the halo-to-halo scatter is due to the fact that during the hierarchical assembly of dark matter haloes, the inner regions of early virialized objects often survive accretion onto a larger system, thus giving rise to a population of subhaloes. Depending on their orbits and their masses, these subhaloes therefore either merge,
are disrupted or survive to the present day. 
These subhalos appear ubiquitous in high resolution cosmological simulations and provide the source of fluctuations.
Fluctuations due to subhalos in parent halos are important for understanding the time evolution of dark matter density profiles and the halo-to-halo scatter of the inner cusp seen in recent ultrahigh resolution cosmological simulations. 
This scatter may be explained by subhalo accretion histories: when we allow for a population of
subhalos of varying concentration and mass, the total inner profile of dark matter can either steepen or flatten.

In the present paper, the single profile that we obtain 
is the average profile, 
the scatter is similar to that observed in N-body simulations. For example, the scatter in the logarithmic gradient at $0.01 r_{vir}$ for a cluster of $M \simeq 10^{14} M_{\odot}$ in case C is $\pm 0.06$, so that the slope is given by $\alpha \simeq 0.98 \pm 0.06$, similarly to Diemand et al. (2004). 

\subsection{Mass dependence of inner slopes in case C} 

In order to have more insights on mass dependence of slope in case C, in the following, we develop, the same analysis of Merrit et al. (2006), but using instead of their N-body simulations, our model to study halo formation and evolution.  
They extracted the density profiles, from a set of $N$-body dark matter halos and compared the $N$-body density profiles
with various parametric models to find which provide the best fit.   
They found on overall that the best-fitting model to the $\Lambda$CDM halos is Einasto's, although the Prugniel-Simien 
model also perform well. From the analysis they find
systematic variation in profile shape with halo mass.

\subsubsection{Einasto's, de Vaucouler's and Sersic's profiles}

Before going on we recall what kind of profiles are the one previously mentioned.
As previously reported, Navarro et al. (2004) argued for a model, like de 
Vaucouleurs', in which the logarithmic slope varies continuously
with radius:
\begin{equation}
\rho(r)\propto \exp\left(-A r^\alpha \right),
\label{eq:Sers}
\end{equation}
where $r_{-2}$ is the radius at which the logarithmic slope of
the density is $-2$ and $\alpha$ is a parameter  describing
the degree of curvature of the profile.
Merrit et al. (2005) pointed out that this is the same relation
between slope and radius that defines Sersic's (1963, 1968) law, with
the difference that Sersic's law is traditionally applied to the
projected (surface) densities of galaxies, not to the space density.
%
%
S\'ersic's law -- the function that is so successful at 
describing the luminosity profiles of early-type galaxies and bulges
(e.g. Caon, Capaccioli, \& D'Onofrio 1993; 
Graham \& Guzm\'an 2003, and references therein),
and the projected density of hot gas in galaxy clusters
(Demarco et al. 2003) -- is also an excellent description of
$N$-body halos.
Eq. (\ref{eq:Sers}) is also known as Einasto's law, since 
Einasto's (1965, 1968, 1969) 
made an extensive use of equation~(\ref{eq:Sers}) to model the 
light and mass distributions of galaxies (see also Einasto \& Haud 1989).
In addition, we henceforth replace the exponent $\alpha$ by $1/n$ in 
keeping with the usage established by S\'ersic and de Vaucouleurs.
%
%
S\'ersic (1963, 1968) generalized de Vaucouleurs' (1948) $R^{1/4}$ 
luminosity profile model 
by replacing the exponent $1/4$ with $1/n$, such that $n$ was a free 
parameter that measured the `shape' of a galaxy's luminosity profile.
Using the observers' notion of `concentration' 
(see the review in Graham, Trujillo, \& Caon 2001), 
the quantity $n$ is monotonically related to how centrally concentrated 
a galaxy's light profile is. 
With $R$ denoting the {\it projected} radius, 
S\'ersic's 
model is often written as 
\begin{equation}
I(R)=I_{\rm e} \exp\left\{ -b_n\left[ (R/R_{\rm e}) ^{1/n} -1\right] \right\},
\label{Sersic}
\end{equation}
where $I_{\rm e}$ is the (projected) intensity at the (projected)
effective radius $R_{\rm e}$.  The term $b_n$ is not a parameter but a
function of $n$ and defined in such a way that $R_{\rm e}$ encloses
half of the (projected) total galaxy light (Caon et al.\ 1993; see
also Ciotti 1991, his equation (1)).  A good approximation when $n \geq 0.5$ is given in Prugniel \& Simien (1997) as 
\begin{equation}
b_n \approx 2n-1/3+0.009876/n.
\label{Eqbutt}
\end{equation}

\subsubsection{Mass dependence of slope}

In order to have some insight on the quoted problem, namely the mass dependence of the slope in case C,
we used a method similar to that of Merrit et al. (2006), who plotted their $N$-body halos, together with real elliptical galaxies and clusters, in the profile shape vs. mass plane.  
In particiluar, they used the profile shape parameter $n$ from the S\'ersic model fit to the light profile, or the corresponding parameter from the Prugniel-Simien model fit to the dark-matter density profiles obtained through Del Popolo (2009). In the present paper, we use the results of Merrit et al. (2006) for what concerns galaxies and clusters, while 
for the dark matter halos calculated with our model, we used the Prugniel-Simien model fit to the dark-matter density.
As in Merrit et al. (2006), in Fig. 4, we plotted real elliptical galaxies and clusters: galaxy clusters from Demarco et al. (2003); dwarf ellipticals (dE) galaxies from Bingeeli \& Jerjen (1998); dE galaxies from Stiavelli et al. (2001); dE galaxies from Graham \& Guzman (2003); bright elliptical galaxies from Caon et al. (19939 and D'Onofrio et al. (1994). We also 
plotted galaxy clusters and galaxies haloes obtained with the method of the present paper for the case C. 
Similarly to Merrit et al. (2006), Fig. 4 suggests that the galaxy-sized halos obtained with our model have a different shape parameter, i.e. a different mass 
distribution, than the cluster-sized halos, obtained with the same model.
The sample of dwarf- and galaxy-sized halos in Fig. 4
has a mean ($\pm$ standard deviation)\footnote{Reminder: the uncertainty on the mean is not equal to the standard deviation.} 
profile shape $n=3.72 (\pm 0.8)$, while the cluster-sized halos had $n=2.85 (\pm 0.6)$.  
The same conclusion was reached by Merritt et al. (2005)
who studied a different sample of $N$-body halos.
The sample of dwarf- and galaxy-sized halos from that paper
had a mean ($\pm$ standard deviation)
profile shape $n=3.04 (\pm 0.34)$, 
while the cluster-sized halos had $n=2.38 (\pm 0.25)$. 
In Merrit et al. (2006), they observe the same systematic difference in their $N$-body halos. 
Taking the profile shape $n$ from the Prugniel-Simien model fits to
the density profile (equivalent to the value of $n$ obtained by 
fitting S\'ersic's 
model to the projected distribution) 
they find $n=3.59 (\pm0.65)$ for their cluster-sized halos and 
$n=2.89 (\pm0.49)$ for their galaxy-sized halos.
A Student $t$ test, without assuming equal
variance in the two distributions, reveals that the means in our model 
are different at the 99.90\% level.
In the Merrit et al. (2006) means are different at the 88\% level. Applying the same test to the data of
Merritt et al. (2005; their Table~1, column 2), which is double the
size of Merrit at al. (2006) sample and also contains dwarf galaxy-sized halos, they found
that the means are different at the 99.98\% level.  
This leads to conclude that there is a significant mass dependence in the
density profiles of the halos in our model similarly to the simulated dark-matter halos in Merrit et al. (2005, 2006).
Density profiles of more massive halos exhibit more curvature
(smaller $n$) on a log-log plot.


\subsubsection{Comparison with N-body simulations}

 An extended description of the limits and advantages of SIM, when compared to N-body simulations, has been 
emphasized in Del Popolo (2009). Here, we just recall that several papers in literature described the reasons why SIM gives accurate results is spite of the fact that it considers collapse and virialization of halos that are spherically symmetric, that have suffered no major mergers, and that have suffered quiescent accretion (e.g., Zaroubi et al. 1996; Naim \& Hoffman 1996; Toth \& Ostriker 1992; Huss et al. 1999a, b; Moore et al. 1999; Ascasibar et al. 2004, 2007). 
In general, the inner slopes and density profiles generated using SIM and our method are different in character from the
profiles predicted by numerical simulations, however the situation changes going from case A to case C.
In case A and B, the``semi-analytic'' halos are more extended, have flatter rotation curves and have higher specific angular momentum (in agreement to Williams et al. 2004 results), in comparison to those formed in N-body simulations.
This result is unsurprising since the types of evolution that numerical N-body and our halos undergo are rather different:
the former are produced as a cumulative result of many minor and major mergers of smaller sub-halos, while the latter are the product of quiescent accretion of lumpy matter onto the primary halo, moreover in case A we take into account the effects of barions.

In case A, the flattening of the inner slopes of haloes is produced by the role of angular momentum, dynamical friction 
and the interplay between dark matter and baryonic component, which is not taken into account by N-body simulations.
As previously reported, our results shows a steepening of the density profile with increasing mass with a density profile of haloes of mass $ > 10^{12} M_{\odot}$ having slopes $>0.8$. 
This is in agreement with recent N-body simulations having a 
logarithmic slope that decreases inward more gradually than the NFW profile (Hayashi et al. 2003; Navarro et al. 2004; Stadel et al. 2008).  
In the case of Stadel et al. (2008) the logarithmic slope is $0.8$ at $0.05 \%$ of $r_v$.


In case B, we do not consider the presence of baryons, so one pertinent question to ask 
is why in case B we obtain different slopes with respect to N-body simulations results. The answer, already given in Del Popolo (2009), is connected to the different predictions for the angular momentum in our model (and also Williams et al. 2004) and N-body simulations. 

As Shown in Fig. 3, the solid histogram representing the total specific angular momentum distribution of the density profile reproducing the NFW halo (described in the previously quoted experiment) is more centrally concentrated than the total specific angular momentum distribution of our reference haloes (dashed histogram), and is closer to those of typical halos emerging from numerical simulations. In Fig. 3, the dotted-dashed and dashed line represents the quoted 
distribution for the halo n. 170 and n. 081, respectively, of van den Bosch et al. (2002). The halo n. 170 resembles most of the specific angular momentum distributions, while the halo n. 081 has the shallowest distribution in their simulations. 
This may suggests, in agreement with Williams et al. (2004), that haloes in N-body simulations lose a considerable amount of angular momentum between 0.1 and 1 $r_{v}$. Since virialization proceeds from inside out, this means that the angular momentum loss takes place during the later stages of the halos' evolution, rather then very early on. This is somehow confirmed by the so called angular momentum catastrophe, namely the fact that dark matter halos generated through gas-dynamical simulations
are too small and have too little angular momentum compared to the halos of real disk galaxies, possibly because it was lost during repeated collisions through dynamical friction or other mechanisms (van den Bosch et al. 2002; Navarro \& Steinmetz 2000). 
The problem can be solved invoking stellar feedback processes (Weil et al. 1998), but part of the angular momentum problem seems due to numerical effects, most likely related to the shock capturing, artificial viscosity used in smoothed particle hydrodynamics (SPH) simulations (Sommer-Larsen \& Dolgov 2001).
%

Another of the reasons why N-body simulations could give unreliable results are connected to two-body relaxation. The processes of relaxation is difficult to quantify, but in the large $N$ limit one expects that the discreteness effects inherent to the N-body technique vanish, so one tries to use as large a number of particles as computationally possible.
Unfortunately in most cosmological simulations the importance of two body interactions does not vanish if one
increases $N$, since structure formation in the cold dark matter (CDM) model occurs hierarchically since there is power
on all scales, so the first objects that form in a simulation always contain only a few particles (Moore et al. 2001),
(Binney \& Knebe 2002) (see Del Popolo 2009, Section 3 for an extended discussion).  

In case C, density profiles and slopes are sonsistent with the NFW model. 

Finally, we have checked that the results of the model of the present paper agree with simulations 
when the model of the present paper is put in the same conditions of simulations, namely 
we did not take into account baryons in our model and we fixed the angular momentum as given in Fig. 3 (dashed line) and for the rest we took into account all the effects of case A.
In Fig. 5, we plot the profiles obtained with our model and those predicted by numerical simulations.
The long-dashed line represents the density profile for haloes of $10^{12} h^{-1} M_{\odot}$.
In order to compare the results of the model with those of N-body simulations, we plotted  
the NFW profile (solid line) for haloes with masses equal to $10^{12} h^{-1} M_{\odot}$, 
and the results for the same mass obtained in the Aquarius Project by  Navarro et al. (2008)
(short--dashed line). The plot shows that the result of the present paper is in good agreement with simulations
and similarly to Navarro et al. (2008) 
density profiles deviate slightly but systematically from the NFW model, and are approximated more accurately by 
the Einasto profile 

Density profiles become monotonically shallower inwards, down to the innermost resolved point, with no indication
that they approach power-law behavior. The innermost slope 
is slightly shallower than -1.
Shallower cusps, such as the $r^{-0.75}$ behavior predicted by the model of Taylor
\& Navarro (2001), cannot yet be excluded.

\subsection{Comparison with other studies}

A comparison of the results of the present paper, concerning the density profile shape and inner slopes, with other studies has somehow already been performed in Del Popolo (2009). 
Before starting the quoted comparison, we summarize the results obtained in the present paper. We have studied how inner slopes of relaxed $\Lambda$CDM halos of galaxies and clusters depend on mass and redshift in three cases.
In the first case, that we labelled as case A, we took into account angular momentum, dynamical friction, baryons, baryons adiabatic contraction. We showed that, in the case of galaxies, the inner slopes strongly depend on mass and redshift while for the case of clusters the mass and redshift dependence of the inner slope is much less evident. In the second case, case B, we did not take into account baryons and dynamical friction and we showed that inner slopes depend less on mass and redshift in comparison with case A, both for galaxies and clusters. Both cases, A and B, lead to conclude against the universality of density profiles of halos. 
In the last case, case C, we took into account only angular momentum whose magnitude was reduced to reproduce the angular momentum of N-body simulations. In this case there is a very slight dependence of slopes on mass and redshift and
the density profiles and slopes have a similar behavior to those obtained in N-body simulations.


The results of case A, agrees with that of Romano-Diaz et al. (2008) who studied the DM cusp evolution using
N-body simulations with and without baryons. Their study shows how baryons tend to reduce the inner slope
because of the heating up of the cusp region via dynamical friction (EZ01) or influx of subhalos into the innermost region of the DM halo.
Our results, are in agreement with EZ01, and El-Zant et al. (2004)
arguing that dynamical friction acting on galaxies moving within the DM background opposes
the effect of adiabatic compression by transferring their orbital energy to the DM, thus heating up and softening the inner profile. Moreover, our density profiles and inner slopes agrees very well with Gentile et al. (2004) rotation curves of some LSB galaxies (Fig. 4 in Del Popolo 2009), and with several studies observing shallower inner slopes than simulations (Flores \& Primack 1994; Moore 1994; Burkert 1995; Kravtsov et al. 1998; Salucci \& Burkert 2000; Borriello \& Salucci 2001; de Blok et al. 2001, 2003; de Blok \& Bosma 2002; de Blok 2003; Spekkens et al. 2005). In the case of clusters, the results agrees with observations finding non-cuspy profiles (e.g., Ettori et al. 2002; Sand et al. 2002, 2004). 

For what concerns the case B, our results agree with all those studies showing how the higher is the value of angular momentum the flatter is the inner slope (e.g., Avila-Reese et al. 1998; Nusser 2001; Hiotelis 2002; Le Delliou \& Henriksen 2003; Williams et al. 2004). 

Case C, as previously described gives inner slopes similar to those obtained in N-body simulations (e.g., Navarro et al. 2004).

A comparison of our results concerning the universality of density profiles, with other previous studies, 
especially dissipationless N-body simulations which do not take account of baryons, is more problematic.

For what concerns case A, there are not other studies dealing with the universality of dark matter halos also containing baryons. Even if we cannot make direct comparison with other studies, our results is in agreement with those studies
that have argued against the quoted universality in DM simulations (Jing \& Suto 2000; Subramanian et al. 2000; Ricotti 2003; Ricotti \& Wilkinson 2004; Cen et al. 2004; Graham et al. 2005; Merrit et al. 2005; Ricotti et al. 2007). 
Our result also agrees with 
observational evidences of non-universality in dark matter density profiles, at galaxy (Simon et al. 2003a,b; 2004) and on cluster scales (Host \& Hansen 2009). 
Similar conclusions are valid for Case B.
For what concerns case C, our result is in agreement with Merrit et al. (2006) concluding that there is a significant mass dependence in the
density profiles of the halos.

Always in case C, to better explore how the homology (i.e., universality) of 
CDM halos is broken, it would be beneficial to analyze a large, low-resolution sample of
halos from a cosmological cube simulation in order to obtain good
statistics. Moreover, the presence of large subhalos, possible debris wakes from larger structures, 
and the collective impact from differing degrees of virialization in the outer
regions
could be quantified.

\section{Conclusions}

In the present paper, we studied the dependence of the inner slope of the density profiles of dark matter halos with and without baryons. 
The density profiles were built up using the Del Popolo (2009) method, which includes the effect of baryons on dark matter halos. 
We calculated the inner slopes for three different cases: in the case A we take into account all the effects included in Del Popolo (2009), namely angular momentum, dynamical friction, baryons, baryons adiabatic contraction; in the case B there are no baryons and dynamical friction, and angular momentum is taken into account as in Del Popolo (2009); in case C only angular momentum is taken into account with magnitude reduced as in Del Popolo (2009) in order to reproduce the angular momentum of N-body simulations. 
In case A, and B, we found a strong dependence of the inner slope with mass and redshift: lower mass objects having smaller slopes which increases with redshift. 
This result argues against the universality of these kind of profiles. 
When baryons are not present (case C), the mass and redshift dependence of slope is quite slight in agreement with results of other studies (e.g., Gao et al. 2008).
However, to have some further insights on the variation of the profile shape with the halo mass,   
we used a method similar to that of Merrit et al. (2006), who plotted their $N$-body halos, in the profile shape vs. mass plane.  
Similarly to Merrit et al. (2006), the galaxy-sized halos obtained with our model have a different shape parameter, i.e. a different mass distribution than the cluster-sized halos obtained with the same model.
The sample of dwarf- and galaxy-sized halos studied has a mean ($\pm$ standard deviation) profile shape $n=3.72 (\pm 0.8)$, while the cluster-sized halos had $n=2.85 (\pm 0.6)$. A Student $t$ test, without assuming equal variance in the two distributions, reveals that the means in our model 
are different at the 99.90\% level. This leads to conclude that there is a significant mass dependence in the
density profiles of the halos in our model similarly to the simulated dark-matter halos in Merrit et al. (2005, 2006).

\section*{Acknowledgments}

I thank Professor B. Moore for some helpful suggestions.

\end{document}